\documentclass[11pt,reqno]{amsart}
\usepackage{geometry} 
\usepackage{tikz}
\usepackage{tikz-cd}
\usepackage{color}
\usepackage{hyperref}

\newcommand{\ed}{\mathrm{d}}        
\newcommand{\mR}{\mathrm{R}}

\title[Strong Cosmic Censorship for Spherically Symmetric Einstein-dust Model]{Note on Strong Cosmic Censorship for Spherically Symmetric Einstein-dust Model}
\author{Pengyu Le}
\address{Yanqi Lake Beijing Institute of Mathematical Sciences and Applications, Beijing, China}
\email{pengyu.le@bimsa.cn}
\date{} 

\begin{document}

\begin{abstract}
In this note, we derive an ordinary differential equation for outgoing light rays in the spacetime of a spherically symmetric contracting dust cloud. The violation of strong cosmic censorship is equivalent to the existence of a solution blowing up at the centre of the symmetry. As an application, we reprove the generic violation of strong cosmic censorship for a dust cloud first proved by Christodoulou. We also derive a similar equation for a spherically symmetric dust cloud in the case of positive cosmological constant and show the generic violation of strong cosmic censorship in this case.
\end{abstract}
\maketitle
\tableofcontents

\section{Introduction}\label{section 1}
\noindent
The equations of Einstein-dust model with cosmological constant $\Lambda$ are
\begin{align}
\left\{
\begin{aligned}
&
\mR_{\mu\nu} - \frac{\mR}{2} g_{\mu\nu} + \Lambda g_{\mu\nu} = 8\pi T_{\mu\nu},
\\
&
T_{\mu\nu} = \rho n_{\mu} n_{\nu},  \quad g_{\mu\nu} n^{\mu} n^{\nu} =-1.
\end{aligned}
\right.
\end{align}
Tolman \cite{T} studied the spherically symmetric case and obtained the solutions. Oppenheimer and Snyder \cite{OS} considered a special solution corresponding to a homogeneous spherically symmetric dust cloud and found that its area radius approaches to the gravitational radius within finite proper time on the surface of the dust. 

Christodoulou \cite{C} investigated the behaviour of outgoing light rays for general inhomogeneous Tolman solutions. He analysed the differential equations for outgoing light rays thoroughly and proved the surprising results: in the generic case strong cosmic censorship is false and for an open subset of initial densities weak cosmic censorship is false as well. A key step is to transform the ordinary differential equation for the outgoing light ray emanating from the singular point at the centre into a form in which the singular point at the centre appears as a regular singular point.

In this note, we shall derive another ordinary differential equation for outgoing light rays. The equation has the feature that its solution parametrises an outgoing light ray emanating from the singular point at the centre if and only if it blows up at the centre. As an application, with the help of this equation, we reprove the violation of strong cosmic censorship in the generic case first proved in \cite{C}. We shall also derive a similar equation for a spherically symmetric dust cloud in the case of positive cosmological constant $\Lambda>0$ and apply the equation to show the generic violation of strong cosmic censorship in this case.

\section{The ordinary differential equation for outgoing light rays}\label{section 2}
\noindent
Following \cite{C}, We describe Tolman's solution corresponding to a spherically symmetric dust cloud which is initially at rest.

Let $\tau$ be the co-moving coordinate of the dust and $R$ label the spherical shells of dust. Let $r(\tau,R)$ be the area radius of the $2$-sphere $\tau= \mathrm{const}, R= \mathrm{const}$. Assume the labelling $R$ be the area radius on the initial surface $\tau=0$, i.e.
\begin{align}
r(0,R) = R.
\end{align}
Let $\rho(R)$ denote the initial mass density and $m(R)$ denote the initial mass distribution:
\begin{align}
m(R) = 4\pi \int_0^R \rho(S) S^2 \ed S.
\end{align}
Let $a(R)$ denote the mean density within the sphere of radius $R$ on the initial surface:
\begin{align}
a(R) = \frac{3m(R)}{4\pi R^3}.
\end{align}

The solution of the Einstein-dust model with zero cosmological constant for a spherically symmetric dust which is initially at rest is given by the following
\begin{align}
\ed S^2 = - \ed \tau^2 + e^{2\omega} \ed R^2 + r^2 \ed \Sigma^2,
\end{align}
where $\ed \Sigma^2$ is the canonical metric of the $2$-sphere and 
\begin{align}
e^{\omega} = \frac{r'}{\left( 1- 2m/R\right)^{\frac{1}{2}}},
\end{align}
where $r(\tau,R)$ is given by the parametric equations
\begin{align}
\label{eqn 7}
\left\{
\begin{aligned}
&
\tau = \left( \frac{3}{32\pi a} \right)^{\frac{1}{2}} (\eta + \sin \eta),
\\
&
r= \frac{R}{2}(1+ \cos \eta).
\end{aligned}
\right.
\end{align}
Let the prime denote the spatial derivate w.r.t. $R$. From the parametric equations \eqref{eqn 7}, $r'$ is given by
\begin{align}
r'= \frac{1}{2} (1+ \cos \eta) - \frac{1}{4} \frac{\sin \eta(\eta+ \sin \eta)}{1+\cos \eta} \frac{R a'}{a}.
\end{align}
The equation of the outgoing light ray is
\begin{align}
\frac{\ed \tau}{\ed R} = e^{\omega} = \frac{r'}{\left( 1- 2m/R\right)^{\frac{1}{2}}}.
\end{align}

Let $\tau_s(R)$ be the parametrisation of the surface of singular points $r=0$ and $\tau_a(R)$ be the parametrisation of the apparent horizon $r=2m(R)$:
\begin{align}
&
\tau_s(R) = \left( \frac{3\pi}{32 a(R)} \right)^{\frac{1}{2}}
\\
&
\tau_a(R)
=
\left( \frac{3}{32 \pi a(R)} \right)^{\frac{1}{2}}
\left[ \cos^{-1} \left( \frac{4m(R)}{R} -1 \right) + \sqrt{ 1- \left( \frac{4m(R)}{R} -1 \right)^2}\right]
\end{align}

Introduce the notations:
\begin{align}
&
\alpha = \rho(0),
\quad
\beta= -\frac{1}{2} \rho''(0), \footnotemark
\\
&
R_1 = \left( \frac{\alpha}{\beta} \right)^{\frac{1}{2}},
\quad
x=R/R_1,
\\
&
\gamma = \frac{8\pi}{3} \frac{\alpha^2}{\beta},
\\
&
\rho_0(x) = \frac{1}{\alpha} \rho (R),
\quad
a_0(x) = \frac{1}{\alpha} a(R),
\\
&
b(x): \quad a_0(x) = 1- \frac{3}{5} x^2 b(x),
\\
&
c(x): \quad a_0'(x) = -\frac{6}{5} x c(x),
\\
&
\tau_0  = \tau_s(0) = \tau_a(0) = \left( \frac{3\pi}{32 \alpha}\right)^{\frac{1}{2}},
\end{align}
and
\begin{align}
&
\zeta= \left( \frac{32\pi \alpha}{3} \right)^{\frac{1}{2}} (\tau -\tau_0) = \frac{\eta+\sin \eta}{a_0(x)^{\frac{1}{2}}} - \pi,
\\
&
\delta = \pi -\eta.
\end{align}
\footnotetext{Throughout this note, we only consider the initial mass density function $\rho(R)$ with nonzero second order derivative $\rho''(0) \neq 0$.}

Christodoulou \cite{C} used $\zeta$ and $x$ as variables and derived the following ordinary differential equation for outgoing light rays
\begin{align}
\label{eqn 24}
\frac{\ed \zeta}{\ed x} 
= 
2\gamma^{\frac{1}{2}} \left( 1 - \gamma x^2 a_0(x) \right)^{-\frac{1}{2}}
\left[ 
(1+ \cos \eta) - \frac{1}{4} \frac{\sin \eta(\eta+ \sin \eta)}{1+\cos \eta} \left(-\frac{6}{5} \frac{x^2 c(x)}{1-\frac{3}{5} x^2 b(x)}\right) 
\right].
\end{align}
He discovered that the equation has an non-negative solution with the form
\begin{align}
\zeta = \frac{3\pi}{10} x^{7/3} \theta,
\end{align}
where $\theta$ is a regular function and $\theta(0) = \lambda = \left( \frac{15}{\pi} \right)^{\frac{1}{3}} \gamma^{1/2}$ at the singular point at the centre.

In the following, we derive an ordinary differential equation for outgoing light rays in terms of other variables $\delta$ and $x$ from equation \eqref{eqn 24}. Introduce the functions $k,p,q$
\begin{align}
&
k(\delta) = \frac{\delta - \sin \delta}{\frac{1}{6} \delta^3},
\\
&
p(\delta) = \frac{1-\cos \delta}{\frac{1}{2} \delta^2},
\\
&
q(\delta) = \frac{\delta\sin \delta}{2(1-\cos \delta)}.
\end{align}
$k(0)=p(0)=q(0)=1$. $\zeta$ expressed in terms of $\delta$ and $x$ is
\begin{align}
\zeta 
&= -\frac{1}{6} \delta^3 k(\delta) \left( a_0(x) \right)^{-\frac{1}{2}} + \frac{3\pi}{10} x^2 d(x)
\end{align}
where $d(x)$ is the function
\begin{align}
d(x) = \frac{10}{3}\frac{\left( a_0(x) \right)^{-\frac{1}{2}} -1}{x^2}.
\end{align}
Therefore, the left hand side of equation \eqref{eqn 24} expressed in terms of $\delta$ and $x$ is
\begin{align}
\frac{\ed \zeta}{\ed x} = 
& - \frac{1}{2} \delta^2 \left( k(\delta) + \frac{1}{3} \delta k'(\delta) \right)  \left( a_0(x) \right)^{-\frac{1}{2}} \frac{\ed \delta}{\ed x}
\nonumber
\\
&
- \frac{1}{10} \delta^3 k(\delta)  \left( a_0(x) \right)^{-\frac{3}{2}} c(x) x+ \frac{3\pi}{5} x \left( d(x) + \frac{1}{2}x d'(x) \right).
\end{align}
And the right hand side of equation \eqref{eqn 24} expressed in $\delta$ and $x$ is
\begin{align}
2\gamma^{\frac{1}{2}} 
\left( 1- \gamma x^2 a_0(x) \right)^{-\frac{1}{2}}
\left[ 
\frac{1}{4} \delta^2 p(\delta)  - \frac{1}{2} \frac{q(\delta)}{\delta} \left(\pi - \frac{1}{6} \delta^3 k(\delta) \right) \left( -\frac{6}{5} \frac{c(x)}{1- \frac{3}{5} x^2 b(x)} \right) x^2 
\right]
\end{align}
Thus substituting both sides to equation \eqref{eqn 24}, we obtain the following equation for outgoing light rays in terms of $\delta$ and $x$
\begin{align}
\frac{\ed \delta}{\ed x}
=&
\frac{6\pi}{5} \frac{\left(d(x) + \frac{1}{2}x d'(x)\right) \left(1- \frac{3}{5} x^2 b(x) \right)^{\frac{1}{2}}}{k(\delta) + \frac{1}{3} \delta k'(\delta)} x \delta^{-2}
\nonumber
\\
&
- \frac{1}{5} \frac{c(x)}{1-\frac{3}{5}x^2 b(x)}  \frac{k(\delta)}{k(\delta) + \frac{1}{3} \delta k'(\delta)} x \delta
\nonumber
\\
&
-\frac{12}{5}  \gamma^{\frac{1}{2}} 
\frac{c(x)}{\left( 1- \gamma x^2 a_0(x) \right)^{\frac{1}{2}} \left(1- \frac{3}{5} x^2 b(x) \right)^{\frac{1}{2}}}
\frac{q(\delta)\left(\pi - \frac{1}{6} \delta^3 k(\delta) \right) }{k(\delta) + \frac{1}{3} \delta k'(\delta)} 
x^2  \delta^{-3}
\nonumber
\\
&
-\gamma^{\frac{1}{2}} 
\left( \frac{1- \frac{3}{5} x^2 b(x)}{1- \gamma x^2 a_0(x)} \right)^{\frac{1}{2}}
\frac{p(\delta)}{k(\delta) + \frac{1}{3} \delta k'(\delta)}.
\label{eqn 33}
\end{align}
The above equation can be rewritten as
\begin{align}
\frac{\ed \delta}{\ed x}
=&
F(\gamma,x,\delta)
\nonumber
\\
=&
A(x,\delta) x \delta^{-2} 
- B(x,\delta)  x\delta 
- \frac{\gamma^{\frac{1}{2}}C(x,\delta)}{\left( 1- \gamma x^2 a_0(x) \right)^{\frac{1}{2}}} x^2 \delta^{-3}
- \frac{\gamma^{\frac{1}{2}}D(x,\delta)}{\left( 1- \gamma x^2 a_0(x) \right)^{\frac{1}{2}}},
\label{eqn 34}
\end{align}
where $A,B,C,D$ are functions
\begin{align}
A(x,\delta)=&
\frac{6\pi}{5} \frac{\left(d(x) + \frac{1}{2}x d'(x)\right) \left(1- \frac{3}{5} x^2 b(x) \right)^{\frac{1}{2}}}{k(\delta) + \frac{1}{3} \delta k'(\delta)},
\quad
A(0,0) = A_0 = \frac{6\pi}{5},
\\
B(x,\delta)=&
\frac{1}{5} \frac{c(x)}{1-\frac{3}{5}x^2 b(x)}  \frac{k(\delta)}{k(\delta) + \frac{1}{3} \delta k'(\delta)},
\quad
B(0,0) = B_0 =\frac{1}{5},
\\
C(x,\delta)=&
\frac{12}{5}
\frac{c(x)}{\left(1- \frac{3}{5} x^2 b(x) \right)^{\frac{1}{2}}}
\frac{q(\delta)\left(\pi - \frac{1}{6} \delta^3 k(\delta) \right) }{k(\delta) + \frac{1}{3} \delta k'(\delta)},
\quad
C(0,0) = C_0 = \frac{12\pi}{5},
\\
D(x,\delta)=&
\left( {1- \frac{3}{5} x^2 b(x)} \right)^{\frac{1}{2}}
\frac{p(\delta)}{k(\delta) + \frac{1}{3} \delta k'(\delta)},
\quad
D(0,0) = D_0 = 1.
\end{align}
Similar to equation \eqref{eqn 24}, if there exists an non-negative solution with the initial data $\delta(0)=0$ for equation \eqref{eqn 34}, then the solution parametrises an outgoing light ray emanating from the singular point at the centre. However equation \eqref{eqn 34} is singular at $\delta=0$, thus the existence of the solution is problematic. We will solve the existence problem of equation \eqref{eqn 34} in the next section. To conclude this section, we transform equation \eqref{eqn 24} by the following change of variables
\begin{align}
\iota = \log \delta,
\end{align}
then obtain
\begin{align}
\frac{\ed \iota}{\ed x}
=&
G(\gamma,x,\iota) = e^{-\iota} F(\gamma, x, e^{\iota})
\nonumber
\\
=&
A(x,e^{\iota}) x e^{-3 \iota} 
- B(x,e^{\iota})  x 
- \frac{\gamma^{\frac{1}{2}}C(x,e^{\iota})}{\left( 1- \gamma x^2 a_0(x) \right)^{\frac{1}{2}}} x^2 e^{-4\iota}
- \frac{\gamma^{\frac{1}{2}}D(x,e^{\iota})}{\left( 1- \gamma x^2 a_0(x) \right)^{\frac{1}{2}}} e^{-\iota}.
\label{eqn 40}
\end{align}
Equation \eqref{eqn 40} is equivalent to equation \eqref{eqn 34} for $\delta>0$. If there exists a solution blowing up to $-\infty$ at $x=0$, then it parametrises an outgoing light ray emanating from the singular point at the centre.

\section{Outgoing light rays from the singular point at the centre}\label{section 3}
\noindent
To prove that equation \eqref{eqn 34} enables a solution with the initial data $\zeta(0)=0$, instead of solving the equation from the singular point, we can solve the equation backward from some regular point and show the solution decreasing to zero when approaching $x=0$. Equivalently, we will solve equation \eqref{eqn 40} backward starting from some positive $x$ and prove that the solution blows up to $-\infty$ at $x=0$.

We first construct a class of comparison functions for equation \eqref{eqn 40}. The comparison function takes the form of
\begin{align}
\iota_c(x) = n \log x + \log w, \quad e^{\iota_c} = w x^{n}, \quad n,w >0,
\end{align}
and satisfies the differential inequality
\begin{align}
\frac{\ed \iota_c}{\ed x} \leq G(\gamma, x, \iota_c).
\end{align}
Substituting the comparison function $\iota_c$ into the differential inequality, we obtain that
\begin{align}
A(x, w x^{n}) w^{-3} x^{1-3n}
\geq&
n x^{-1} 
+B(x, w x^{n})  x 
+\frac{\gamma^{\frac{1}{2}}C(x,w x^{n})}{\left( 1- \gamma x^2 a_0(x) \right)^{\frac{1}{2}}} w^{-4} x^{2-4n}
\nonumber
\\
&
+\frac{\gamma^{\frac{1}{2}}D(x,w x^{n})}{\left( 1- \gamma x^2 a_0(x) \right)^{\frac{1}{2}}} w^{-1} x^{-n}.
\label{eqn 43}
\end{align}
The above inequality holds for some small interval $(0,x_{\iota_c}]$ depending on $n,w$ if $n$ satisfies the following inequalities
\begin{align}
1-3n <-1, \quad 1-3n < 1, \quad 1-3n <2-4n, \quad 1-3n <-n,
\end{align}
which have an nonempty solution set
\begin{align}
\frac{2}{3} < n < 1.
\end{align}
For the case $n=1$, if $w$ satisfies
\begin{align}
A_0 w^{-3} > \gamma^{\frac{1}{2}} C_0 w^{-4}  
\quad
\Leftrightarrow
\quad
w > \gamma^{\frac{1}{2}} C_0/A_0 
\quad
\Leftrightarrow
\quad
w > 2\gamma^{\frac{1}{2}},
\end{align}
then there exists a small interval $(0,x_{\iota_c}]$ depending on $w$ such that differential inequality \eqref{eqn 43} holds. Similarly for $n=\frac{2}{3}$, if $w$ satisfies
\begin{align}
A_0 w^{-3} > \frac{2}{3}
\quad
\Leftrightarrow
\quad
0< w < \left( \frac{3A_0}{2} \right)^{\frac{1}{3}}
\quad
\Leftrightarrow
\quad
0< w < \left( \frac{9\pi}{5} \right)^{\frac{1}{3}},
\end{align}
then there exists a small interval $(0,x_{\iota_c}]$ depending on $w$ such that inequality \eqref{eqn 43} holds.

Suppose that $\iota_c$ is one of the comparison functions constructed above such that differential inequality \eqref{eqn 43} holds on $(0,x_{\iota_c}]$. Then by the comparison theorem of solutions for ordinary differential equations, we obtain that if a solution $\iota(x)$ of equation \eqref{eqn 40} is bounded from above by the comparison function $\iota_c(x)$ at some point $x_0 \in (0,x_{\iota_c}]$, i.e.
\begin{align}
\iota(x_0) \leq \iota_c(x_0),
\end{align}
then $\iota(x)$ is bounded from above by $\iota_c(x)$ for all $x\in (0,x_0]$ whenever $\iota(x)$ is finite,
\begin{align}
\iota(x) \leq \iota_c(x), \quad \forall x \in (0,x_0] \cap \{x: \iota(x) > -\infty\}.
\end{align}
Since the comparison function $\iota_c(x)$ blows up to $-\infty$ at $x=0$, then the solution $\iota(x)$ must blow up either at $x=0$ or at some positive $x$. Recall that $\iota(x) = -\infty$ is equivalent to $\delta(x)=0$ and $r=0$, which means that an outgoing light ray emanates from the corresponding point $(\tau, R)$ on the singular surface. However it was showed in \cite{C} that there exists no outgoing light ray emanating from singular surface except from the singular point at the centre, thus $\iota(x)$ must blow up at $x=0$.

Therefore with the help of the comparison functions, we prove there exists an infinity of solutions for equation \eqref{eqn 40} which blow up to $-\infty$ at $x=0$. It implies that there exists an infinity of outgoing ligh rays emanating from the singular point at the centre. Thus the strong cosmic censorship is false.

\section{Parametric equations in the case $\Lambda>0$}\label{section 4}
\noindent
In the rest of this note, we apply the same method to study outgoing light rays in the collapse of a dust cloud with the presence of a positive cosmological constant $\Lambda>0$. The general Tolman solution is given by the following formulae
\begin{align}
\begin{aligned}
&
\ed S^2 = -\ed \tau^2 + \frac{r'^2}{f^2(R)} \ed R^2 + r^2 \ed \Sigma^2,
\\
&
\int \frac{\ed r}{\sqrt{f^2(R) -1 + \frac{1}{2}F(r)r^{-1} + \frac{\Lambda}{3} r^2}} \pm \tau = \mathbf{F}(R).
\end{aligned}
\end{align}

We consider a class of Tolman's solution where $f^2(R)-1=0$ in the case of positive cosmological constant $\Lambda>0$. Adopt the same coordinate system as in \cite{C} where $R=r(0,R)$ on the initial surface $\tau=0$ and the same notations of the initial mass density $\rho(R)$, the initial mass distribution $m(R)$ and the mean density $a(R)$. $r(\tau,R)$ is given by the following parametric equations
\begin{align}
\left\{
\begin{aligned}
&
\tau=- \frac{2}{\sqrt{3\Lambda}} ( \eta -\eta_0 (R) ),
\\
&
r= R \left( \frac{8\pi a(R)}{\Lambda} \right)^{\frac{1}{3}} \sinh^{\frac{2}{3}} \eta = R \frac{\sinh^{\frac{2}{3}}\eta}{\sinh^{\frac{2}{3}} \eta_0(R) }.
\end{aligned}
\right.
\label{eqn 51}
\end{align}
From parametric equations \eqref{eqn 51}, we obtain the differential of $r$
\begin{align}
\left\{
\begin{aligned}
&
\dot{r}
=
- \sqrt{\frac{\Lambda}{3}} r \coth \eta
=
- \left( \frac{\Lambda r^2}{3} + \frac{2m}{r} \right)^{\frac{1}{2}},
\\
&
r'
=
\frac{r}{R} + \frac{2r}{3} \left( \coth \eta - \coth \eta_0 \right) \eta_0'
\end{aligned}
\right.
\end{align}
The equation of outgoing light rays is
\begin{align}
\frac{\ed \tau}{\ed R} = r' = \frac{r}{R} + \frac{2r}{3} \left( \coth \eta - \coth \eta_0 \right) \eta_0'
\end{align}
and the rate of change for $r$ along outgoing light rays is
\begin{align}
\frac{\ed r}{\ed \tau} = \dot{r}+1 = 1- \left( \frac{\Lambda r^2}{3} + \frac{2m}{r} \right)^{\frac{1}{2}}.
\end{align}

On the initial surface $\tau=0$, we assume that $\frac{\ed r}{\ed \tau}>0$ inside the dust cloud, i.e.
\begin{align}
\frac{\Lambda R^2}{3} + \frac{2m(R)}{R} < 1.
\end{align}
Assume that the equation $\frac{\Lambda r^2}{3} + \frac{2m(R)}{r} =1$ always has two distinct roots $r_a(R)$ and $r_c(R)$. We assume that initially inside the dust cloud, the radius $r$ lies between them, i.e.
\begin{align}
r_e(R) < r(0,R) = R < r_c(R).
\end{align}

From parametric equations \eqref{eqn 51}, we observe that $r$ decreases in time and becomes zero within a finite time period. The singular surface is parametrised by the equation
\begin{align}
\tau = \tau_s(R) = \frac{2}{\sqrt{3 \Lambda}} \eta_0(R).
\end{align}
The apparent horizon is parametrised by the equation
\begin{align}
r_a(R) = r(\tau_a(R), R) =  R \left( \frac{8\pi a(R) }{\Lambda} \right)^{\frac{1}{3}} \sinh^{\frac{2}{3}} \left( \eta_0(R) -\frac{2\tau_a(R)}{\sqrt{3\Lambda}} \right).
\end{align}

The apparent horizon and the singular surface have a contact point at the centre. Since beyond the apparent horizon, the radius $r$ decreases along outgoing light rays, therefore no outgoing light ray can emanate from the singular surface except from the singular point at the centre.

\section{Violation of strong cosmic censorship in the case $\Lambda>0$}\label{section 5}
\noindent
We adapt the calculations in sections \ref{section 2} and \ref{section 3} to the class of solutions discussed in section \ref{section 4}. The equation for outgoing light rays in terms of $\eta$ is
\begin{align}
\frac{\ed \eta}{\ed R}
=
- \frac{\sqrt{3\Lambda}}{2} \frac{\sinh^{\frac{2}{3}} \eta}{\sinh^{\frac{2}{3}} \eta_0}
- \sqrt{\frac{\Lambda}{3}} R \frac{\sinh^{\frac{2}{3}} \eta}{\sinh^{\frac{2}{3}} \eta_0}  \left( \coth \eta - \coth \eta_0 \right) \frac{\ed \eta_0}{\ed R}
+ \frac{\ed \eta_0}{\ed R}
\end{align}

Adopting the notations $\alpha$, $\beta$, $R_1$, $\gamma$, $x$, $\rho_0(x)$, $a_0(x)$, $b(x)$, $c(x)$ the same as in section \ref{section 2}. Then
\begin{align}
\frac{\ed}{\ed R} = \frac{1}{R_1} \frac{\ed}{\ed x}, \quad R\frac{\ed}{\ed R} = x \frac{\ed}{\ed x}.
\end{align}
Furthermore introduce the dimensionless quantity $\kappa$
\begin{align}
\kappa = \frac{1}{8\pi} \frac{\Lambda}{\alpha},
\end{align}
then $\sqrt{\frac{\Lambda}{3}} R_1 = \kappa^{\frac{1}{2}} \gamma^{\frac{1}{2}}$. Express the equation for outgoing light rays in terms of $\eta$ and $x$
\begin{align}
\frac{\ed \eta}{\ed x} 
=
-
\frac{3}{2}\kappa^{\frac{1}{2}} \gamma^{\frac{1}{2}}  \frac{\sinh^{\frac{2}{3}} \eta}{\sinh^{\frac{2}{3}} \eta_0}
-
\kappa^{\frac{1}{2}} \gamma^{\frac{1}{2}} \frac{\sinh^{\frac{2}{3}} \eta}{\sinh^{\frac{2}{3}} \eta_0}
\left( \coth \eta - \coth \eta_0 \right) x \frac{\ed \eta_0}{\ed x}
+
\frac{\ed \eta_0}{\ed x}.
\end{align}

Introduce the following change of variables
\begin{align}
\iota = \frac{1}{3} \log \sinh \eta, 
\quad
e^{3\iota} = \sinh \eta, 
\quad
\iota_0 = \frac{1}{3}\log \sinh \eta_0,
\quad
e^{3\iota_0} = \sinh \eta_0,
\end{align}
then
\begin{align}
\frac{\ed \iota}{\ed x} 
=
\frac{1}{3} \coth \eta \frac{\ed \eta}{\ed x}
= 
\frac{1}{3} e^{-3\iota} \left(1+ e^{6\iota} \right)^{\frac{1}{2}} \frac{\ed \eta}{\ed x}.
\end{align}
Therefore the equation for outgoing light rays in terms of $\iota$ and $x$ is
\begin{align}
\frac{\ed \iota}{\ed x}
=&
\kappa^{\frac{1}{2}} \gamma^{\frac{1}{2}} \left( -\frac{1}{2} + x \frac{\ed \iota_0}{\ed x} \right) e^{-2\iota_0} \left( 1 + e^{6 \iota} \right)^{\frac{1}{2}} e^{-\iota}
\nonumber
\\
& 
+  \frac{\ed \iota_0}{\ed x} \left( 1+e^{6\iota_0} \right)^{-\frac{1}{2}} e^{3\iota_0} \left( 1 + e^{6 \iota} \right)^{\frac{1}{2}} e^{-3\iota}
\nonumber
\\
&
-
\kappa^{\frac{1}{2}} \gamma^{\frac{1}{2}}  x \frac{\ed \iota_0}{\ed x} \left( 1+ e^{6\iota_0} \right)^{-\frac{1}{2}} e^{\iota_0} \left( 1+ e^{6\iota} \right) e^{-4\iota},
\end{align}
where $\iota_0$ is given by parametric equations \eqref{eqn 51},
\begin{align}
e^{3\iota_0} = \left( \frac{\kappa}{a_0(x)} \right)^{\frac{1}{2}},
\quad
\iota_0 = \frac{1}{6} \log \kappa  -\frac{1}{6} \log a_0(x),
\quad
\frac{\ed \iota_0}{\ed x} = -\frac{1}{6} \frac{a_0'}{a_0}.
\end{align}
Substituting $\iota_0$, we obtain the equation for outgoing light rays
\begin{align}
\frac{\ed \iota}{\ed x}
=&
-\frac{1}{2} \kappa^{\frac{1}{6}} \gamma^{\frac{1}{2}} \left( 1 - \frac{2}{5} \frac{c(x)}{a_0(x)} x^2 \right) \left(a_0(x)\right)^{\frac{1}{3}} \left( 1 + e^{6 \iota} \right)^{\frac{1}{2}} e^{-\iota}
\nonumber
\\
& 
- \frac{1}{5} \kappa^{\frac{1}{2}} c(x) a_0(x)^{-\frac{3}{2}} \left( 1+ \frac{\kappa}{a_0(x)} \right)^{-\frac{1}{2}} \left( 1 + e^{6 \iota} \right)^{\frac{1}{2}} x e^{-3\iota}
\nonumber
\\
&
+
\frac{1}{5} \kappa^{\frac{2}{3}} \gamma^{\frac{1}{2}}  c(x) \left(a_0(x)\right)^{-\frac{7}{6}} \left( 1+ \frac{\kappa}{a_0(x)} \right)^{-\frac{1}{2}} \left( 1+ e^{6\iota} \right) x^2 e^{-4\iota}.
\label{eqn 67}
\end{align}
Furthermore equation \eqref{eqn 67} can be rewritten as
\begin{align}
\frac{\ed \iota}{\ed x}
=&
G(\kappa, \gamma, x, \iota)
\nonumber
\\
=&
-
\kappa^{\frac{1}{6}} \gamma^{\frac{1}{2}} A(x,\iota) e^{-\iota}
-
\frac{\kappa^{\frac{1}{2}} B(x,\iota)}{\left(1 + \frac{\kappa}{a_0(x)} \right)^{\frac{1}{2}}} x e^{-3\iota}
+
\frac{\kappa^{\frac{2}{3}} \gamma^{\frac{1}{2}} C(x,\iota)}{\left(1 + \frac{\kappa}{a_0(x)} \right)^{\frac{1}{2}}}  x^2 e^{-4\iota},
\label{eqn 68}
\end{align}
where $A(x,\iota), B(x, \iota), C(x, \iota)$ are functions
\begin{align}
A(x,\iota)
=&
\frac{1}{2} \left( 1 - \frac{2}{5} \frac{c(x)}{a_0(x)} x^2 \right) \left(a_0(x)\right)^{\frac{1}{3}} \left( 1 + e^{6 \iota} \right)^{\frac{1}{2}},
\quad
A(0,0) = A_0 = \frac{1}{2},
\\
B(x,\iota)
=& 
\frac{1}{5} c(x) a_0(x)^{-\frac{3}{2}} \left( 1 + e^{6 \iota} \right)^{\frac{1}{2}},
\quad
B(0,0) = B_0 = \frac{1}{5},
\\
C(x,\iota)
=&
\frac{1}{5} c(x) \left(a_0(x)\right)^{-\frac{7}{6}} \left( 1+ e^{6\iota} \right),
\quad
C(0,0) = C_0 = \frac{1}{5}.
\end{align}

The aim is to find a solution of equation \eqref{eqn 68} which blows up to $-\infty$ at the centre $x=0$. First since no outgoing light ray emanates from the singular surface except from the singular point at centre, we conclude that no solution of equation \eqref{eqn 68} blows up to $-\infty$ from the right except at $x=0$. The rest of the argument is parallel to the one in section \ref{section 3}. Consider the comparison functions $\iota_c$ of the form
\begin{align}
\iota_c(x) = n \log x + \log w, \quad e^{\iota_c} = w x^n, \quad n,w>0,
\end{align}
and satisfies the differential inequality
\begin{align}
\frac{\ed \iota_c}{\ed x} \leq G(\kappa, \gamma, x, \iota_c).
\label{eqn 73}
\end{align}
Substituting $\iota_c$ into the above differential inequality, we obtain that
\begin{align}
\frac{\kappa^{\frac{2}{3}} \gamma^{\frac{1}{2}} C(x, w x^n)}{\left(1 + \frac{\kappa}{a_0(w x^n)} \right)^{\frac{1}{2}}} w^{-4} x^{2-4n}
\geq&
\kappa^{\frac{1}{6}} \gamma^{\frac{1}{2}} A(x, w x^n) w^{-1} x^{-n}
+\frac{\kappa^{\frac{1}{2}} B(x, w x^n)}{\left(1 + \frac{\kappa}{ a_0(w x^n)} \right)^{\frac{1}{2}}} w^{-3} x^{1-3n}
\nonumber
\\
&
+n x^{-1}.
\label{eqn 74}
\end{align}
The above inequality holds for some small interval $(0, x_{\iota_c}]$ depending on $n,w$ if $n$ satisfies
\begin{align}
2-4n < -n,
\quad
2-4n < 1-3n,
\quad
2-4n <-1
\quad
\Rightarrow
\quad
n>1.
\end{align}
For the case $n=1$, the inequality \eqref{eqn 74} holds for some small interval $(0, x_{\iota_c}]$ depending on $w$ if $w$ satisfies
\begin{align}
\kappa^{\frac{2}{3}} \gamma^{\frac{1}{2}} C_0 w^{-4}
>
\kappa^{\frac{1}{2}} B_0 w^{-3}
\quad
\Leftrightarrow
\quad
0< w < \kappa^{\frac{1}{6}} \gamma^{\frac{1}{2}}C_0/B_0
\quad
\Leftrightarrow
\quad
0< w < \kappa^{\frac{1}{6}} \gamma^{\frac{1}{2}}.
\end{align}

Thus with the help of the above constructed comparison functions, following the same argument in section \ref{section 3}, we prove that there exists an infinity of solutions for equation \eqref{eqn 68} which blow up to $-\infty$ at $x=0$. It implies that there exists an infinity of outgoing ligh rays emanating from the singular point at the centre for the class of solutions constructed in section \ref{section 4}. Therefore we show the violation of strong cosmic censorship for this class of solutions, where the second derivative of the initial mass density at the centre $\rho''(0) \neq 0$.

\end{document}